\documentclass[11pt,english,reqno]{amsart}
\usepackage{geometry}

\usepackage{amsthm}
\usepackage{amssymb}
\usepackage{setspace}
\usepackage{tikz}
\usepackage{hyperref}
\usepackage{enumitem}
\usepackage{bbm}
\usetikzlibrary{arrows}
\usetikzlibrary{patterns}
\tikzstyle{vertex}=[draw, circle, scale=.7, fill=white]
\makeatletter
\numberwithin{equation}{section}
\theoremstyle{plain}
\newtheorem{theorem}{Theorem}
  \newtheorem{cor}[theorem]{Corollary}
  \newtheorem{lemma}[theorem]{Lemma}
  \newtheorem{prop}[theorem]{Proposition}
  \newtheorem{obs}[theorem]{Observation}
  \theoremstyle{definition}

  \theoremstyle{plain}
  \theoremstyle{remark}

\newcommand{\Z}{\mathbb{Z}}

\newcommand{\R}{\mathbb{R}}
\newcommand{\cal}[1]{\mathcal{#1}}

\newcommand{\tr}{\operatorname{tr}}
\newcommand{\F}{\mathbb{F}}

\newcommand{\subbound}{\ham_f}
\newcommand{\ham}{\mathcal{H}}
\newcommand{\Span}{\operatorname{span}}
\newcommand{\rank}{\operatorname{rank}}
\newcommand{\fit}{\mathcal{M}}

\newcommand{\ol}[1]{\overline{ #1 }}
\newcommand{\uni}{\mathcal{G}}
\newcommand{\cha}{\operatorname{char}}
\renewcommand{\theta}{\vartheta}
\renewcommand{\hat}[1]{\widehat{#1}}
\renewcommand{\bar}[1]{\mkern 1.5mu\overline{\mkern-1.5mu#1\mkern-1.5mu}\mkern 1.5mu}
\renewcommand{\bf}[1]{\mathbf{#1}}

\newif\ifBCrunhead
\BCrunheadfalse
\makeatletter
\def\hackysuperscript#1{\ifBCrunhead\else \textsuperscript{#1}\fi}
\makeatother

\begin{document}
\title{On a fractional version of Haemers' bound}
\author{Boris Bukh\hackysuperscript{1}}\thanks{1. Carnegie Mellon University, Pittsburgh, PA, USA. \texttt{bbukh@math.cmu.edu}. Supported in part by Sloan Research Fellowship and by U.S.\ taxpayers through NSF CAREER grant DMS-1555149.}
\author{Christopher Cox\hackysuperscript{2}}\thanks{2. Carnegie Mellon University, Pittsburgh, PA, USA. \texttt{cocox@andrew.cmu.edu}. Supported in part by U.S.\ taxpayers through NSF CAREER grant DMS-1555149.}

\begin{abstract}
In this note, we present a fractional version of Haemers' bound on the Shannon capacity of a graph, which is originally due to Blasiak. This bound is a common strengthening of both Haemers' bound and the fractional chromatic number of a graph. We show that this fractional version outperforms any bound on the Shannon capacity that could be attained through Haemers' bound. We show also that this bound is multiplicative, unlike Haemers' bound.
\end{abstract}

\maketitle
\BCrunheadtrue

\section{Introduction}

For graphs $G_1,\dots,G_n$, the \emph{strong product of $G_1,\dots,G_n$}, denoted $G_1\boxtimes\dots\boxtimes G_n$, is the graph on vertex set $V(G_1)\times\dots\times V(G_n)$ where $(v_1,\dots,v_n)\sim(u_1,\dots,u_n)$ if and only if for every $i\in[n]$, either $v_i=u_i$ or $v_iu_i\in E(G_i)$. For brevity, we write $G^{\boxtimes n}=\underbrace{G\boxtimes\dots\boxtimes G}_n$.

The \emph{Shannon capacity} of a graph $G$, introduced by Shannon in~\cite{shannon1956zero}, is
\[
\Theta(G):=\sup_n\alpha(G^{\boxtimes n})^{1/n}=\lim_{n\to\infty}\alpha(G^{\boxtimes n})^{1/n},
\]
where $\alpha(G)$ denotes the independence number of $G$. 
Despite the fact that Shannon defined this parameter in 1956, very little is known about it in general. For example, $\Theta(C_7)$ is still unknown. 

There are two general upper bounds on $\Theta(G)$. Firstly, the theta function, $\theta(G)$, is a bound on $\Theta(G)$ which is the solution to a semi-definite program dealing with arrangements of vectors associated with $G$. Introduced by Lov\'asz in~\cite{lovasz1979shannon}, the theta function was used to verify that $\Theta(C_5)=\theta(C_5)=\sqrt{5}$. Secondly, Haemers' bound, $\ham(G;\F)$, is a bound on $\Theta(G)$ which considers the rank of particular matrices over the field $\F$ associated with the graph $G$. Introduced by Haemers in~\cite{haemers1978upper,haemers1979some}, $\ham(G;\F)$ was used to provide negative answers to three questions put forward by Lov\'asz in~\cite{lovasz1979shannon}.

%

In this paper, we present a strengthening of Haemers' bound by defining a parameter $\subbound(G;\F)$, to which we refer to as the \emph{fractional Haemers bound}. 
After we wrote the paper, we learned from Ron Holzman, that this parameter previously appeared in a thesis of Anna Blasiak \cite[Section 2.2]{blasiak}.
We defer the definition of this parameter to Section~\ref{sec:defn}. We show the following results:
\begin{theorem}[First proved by Blasiak \cite{blasiak}] \label{thm:shannonbound}
For any graph $G$ and a field $\F$, 
\[
\Theta(G)\leq\subbound(G;\F)\leq\ham(G;\F).
\]
\end{theorem}

The following results are new.
\begin{theorem}\label{thm:itsbetter}
For any field $\F$ of nonzero characteristic, there exists an explicit graph $G=G(\F)$ with
\[
\subbound(G;\F)<\min\{\ham(G;\F'),\theta(G)\},
\]
for every field $\F'$.
\end{theorem}
Therefore, $\subbound$ is a strict improvement over both $\ham$ and $\theta$ for some graphs.

\textbf{Remark.} Recently, Hu, Tamo and Shayevitz in~\cite{hts18} constructed a different generalization of $\ham(G;\F)$ using linear programming.
Their bound, $\operatorname{minrk}_\F^*(G)$, satisfies $\subbound(G;\F)\leq\operatorname{minrk}_\F^*(G)$ for every graph $G$ and every field $\F$.
They show that there exists a graph $G$ for which $\operatorname{minrk}_\R^*(G)<\min\{\ham(G;\F'),\theta(G)\}$ for every field $\F'$, so Theorem~\ref{thm:itsbetter} holds also when $\F=\R$.

Recall that Lov\'asz showed $\theta(G\boxtimes H)=\theta(G)\cdot\theta(H)$ for any graphs $G,H$; the fractional Haemers bound shares this property.
\begin{theorem}\label{thm:tensorize}
For graphs $G,H$ and a field $\F$,
\[
\subbound(G\boxtimes H;\F)=\subbound(G;\F)\cdot\subbound(H;\F).
\]
\end{theorem}
This is in contrast to $\ham(G;\F)$. As we will show in Proposition~\ref{prop:hamnotens}, for any field $\F$, $\ham(C_5;\F)\geq 3$, yet $\ham(C_5^{\boxtimes 2};\F)\leq 8$.

Sadly, $\subbound$ does not improve upon the known bounds for the Shannon capacity of odd cycles.
\begin{prop}\label{prop:cycles}
For any positive integer $k$ and any field $\F$, $\subbound(C_{2k+1};\F)=k+{1\over 2}$.
\end{prop}

The organization of this paper is as follows: in Section~\ref{sec:defn}, we will define $\subbound(G;\F)$; in fact, we will provide four equivalent definitions, each of which will be useful. In Section~\ref{sec:proof1}, we will prove Theorems~\ref{thm:shannonbound} and~\ref{thm:tensorize} and Proposition~\ref{prop:cycles}. In Section~\ref{sec:proof2}, we will prove Theorem~\ref{thm:itsbetter} and also show that $\subbound(G;\F)$ and $\subbound(G;\F')$ can differ when $\F$ and $\F'$ are different fields.  We will then briefly look at two attempts to ``fractionalize'' the Lov\'asz theta function in Section~\ref{sec:lovasz}.
We conclude with a list of open problems in Section~\ref{sec:conclusion}
\section{The fractional Haemers bound}\label{sec:defn}.

For a graph $G$ and a field $\F$, a matrix $M=(m_{uv})\in\F^{V\times V}$ is said to \emph{fit} $G$ if $m_{vv}=1$ for all $v\in V$ and $m_{uv}=m_{vu}=0$ whenever $uv\notin E$. Define $\fit_\F(G)$ to be the set of all matrices over $\F$ that fit $G$. The \emph{Haemers bound}~\cite{haemers1978upper,haemers1979some} of $G$ is then defined as
\[
\ham(G;\F):=\min\{\rank(M):M\in \fit_\F(G)\}.
\]
Haemers introduced $\ham(G;\F)$ as an upper bound on $\Theta(G)$ in order to provide negative answers to three questions put forward by Lov\'asz in~\cite{lovasz1979shannon}. When the field $\F$ is understood or arbitrary, we will condense the notation and write $\ham(G)=\ham(G;\F)$.

A drawback of Haemers' bound is that it is always an integer. To combat this, we introduce a fractional version.

Let $\F$ be a field, $d$ be a positive integer and $G$ be a graph, and consider matrices $M$ over $\F$ whose rows and columns are indexed by $V\times[d]$. We can consider $M$ as a block matrix where for $u,v\in V$, the $uv$ block, $M_{uv}$, consists of the entries with indices $(u,i),(v,j)$ for all $i,j\in[d]$. We say that $M$ is a \emph{$d$-representation of $G$ over $\F$} if
\begin{enumerate}
\item $M_{vv}=I_d$ for all $v\in V$, where $I_d$ is the $d\times d$ identity matrix, and
\item $M_{uv}=M_{vu}=O_d$ for all $uv\notin E$, where $O_d$ is the $d\times d$ zero matrix.
\end{enumerate}
Define $\fit_\F^d(G)$ to be the set of all $d$-representations of $G$ over $\F$.
We then define the \emph{fractional Haemers bound} to be
\[
\subbound(G;\F):=\inf\bigg\{{\rank(M)\over d}:M\in\fit_\F^d(G),d\in\Z^+\bigg\}.
\]
Notice that $\fit_\F(G)=\fit_\F^1(G)$, so $\subbound(G;\F)\leq\ham(G;\F)$. Again, when the field is understood or arbitrary, we will condense the notation and simply write $\subbound(G)$. More specifically, if we were to write, e.g., $\subbound(G)\leq\subbound(H)$, it is assumed that the field is the same in both instances.

\subsection{Alternative formulations}

We now set out three equivalent ways to define $\subbound(G)$, each of which will be useful going forward.

For positive integers $d\leq n$ (unrelated to the graph $G$), consider assigning to each $v\in V$ a pair of matrices $(A_v,B_v)\in(\F^{n\times d})^2$. We say that such an assignment is an \emph{$(n,d)$-representation of $G$ over $\F$} if
\begin{enumerate}
\item $A_v^TB_v=I_d$ for every $v\in V$, and
\item $A_u^TB_v=A_v^TB_u=O_d$ whenever $uv\notin E$.
\end{enumerate}
\begin{prop}
For a graph $G$ and a field $\F$,
\[
\subbound(G;\F)=\inf_{n,d}\bigg\{{n\over d}:\text{$G$ has an $(n,d)$-representation over $\F$}\bigg\}.
\]
\end{prop}
\begin{proof}
A matrix $M\in\F^{(V\times[d])\times(V\times[d])}$ has $\rank(M)\leq n$ if and only if it is of the form $M=A^TB$ where $A,B\in\F^{n\times(V\times[d])}$. Let $A_v$ be the submatrix of $A$ consisting of all entries with indices $(i,(v,j))$ for $i\in[n]$ and $j\in[d]$, and let $B_v$ be defined similarly for $B$. With this, for any $u,v\in V$, $A_u^TB_v=M_{uv}$. Therefore, $M\in\fit_\F^d(G)$ if and only if $A_v^TB_v=I_d$ for every $v\in V$ and $A_u^TB_v=A_v^TB_u=O_d$ whenever $uv\notin E$.
\end{proof}

A second way to understand $\subbound(G)$ is by considering the lexicographic product, $G\ltimes H$, which is formed by ``blowing up'' each vertex of $G$ into a copy of $H$. More formally, $V(G\ltimes H)=V(G)\times V(H)$ and $(u,x)\sim(v,y)$ in $G\ltimes H$ whenever either $uv\in E(G)$ or $u=v$ and $xy\in E(H)$. In this context, it easy to verify that $\fit_\F^d(G)=\fit_\F(G\ltimes\ol{K_d})$, so:
\begin{prop}
For a graph $G$,
\[
\subbound(G)=\inf_d{\ham(G\ltimes\ol{K_d})\over d}.
\]
\end{prop}

The last equivalent formulation of $\subbound(G)$ is, in some sense, the most general. Consider matrices $M$ over $\F$ whose rows and columns are indexed by $\{(v,i):v\in V,i\in[d_v]\}$ where $d_v$ is some positive integer assigned to $v$. As with $d$-representations, we consider $M$ as a block matrix where for $u,v\in V$, the $uv$ block, $M_{uv}$, consists of those entries with indices $(u,i),(v,j)$ for all $i\in[d_u],j\in[d_v]$. We say that $M$ is a \emph{rank-$r$-representation of $G$ over $\F$} if
\begin{enumerate}
\item $\rank(M_{vv})\geq r$ for all $v\in V$, and
\item $M_{uv}=M_{vu}^T=O_{d_u\times d_v}$ whenever $uv\notin E$, where $O_{d_u\times d_v}$ is the $d_u\times d_v$ zero matrix.
\end{enumerate}
\begin{prop}
For a graph $G$ and a field $\F$,
\[
\subbound(G;\F)=\inf_{M,r}\bigg\{{\rank(M)\over r}:\text{$M$ is a rank-$r$-representation of $G$ over $\F$}\bigg\}.
\]
\end{prop}
\begin{proof}
The lower bound is immediate as an $r$-representation of $G$ is also a rank-$r$-representation.

For the other direction, let $M$ be a rank-$r$-representation of $G$ over $\F$ for some $r$. As $\rank(M_{vv})\geq r$ for all $v\in V$, we can find an $r\times r$ submatrix of $M_{vv}$ of full rank, call this submatrix $M_{vv}'$. Let $M'$ be the submatrix of $M$ induced by the blocks $\{M_{vv}':v\in V\}$; we index the rows and columns of $M'$ by $V\times[r]$. 
For any fixed $v\in V$, $M_{vv}'$ has full rank, so we may perform row operations on $M'$ using only the rows indexed by $\{v\}\times[r]$ to transform $M_{vv}'$ into $I_r$. Let $M''$ be the matrix formed by doing this for every $v\in V$. Therefore, $M''_{vv}=I_r$ for every $v\in V$. Further, as all row operations occurred only between rows corresponding to the same vertex, if $M_{uv}=O_{d_u\times d_v}$, then we also have $M''_{uv}=O_r$. Thus, as $M$ was a rank-$r$-representation of $G$, $M''$ is an $r$-representation of $G$. We conclude that
\[
\subbound(G)\leq{\rank(M'')\over r}={\rank(M')\over r}\leq{\rank(M)\over r},
\]
so the same is true of the infimum over all $M$ and $r$.
\end{proof}

\textbf{Remark.} 
While this paper was in submission, Lex Schrijver introduced us to the following equivalent definition of $\subbound(G;\F)$.
For fixed positive integers $d\leq n$, a collection of subspaces $\{S_v\leq \F^n:v\in V\}$ is called an $(n,d)$-subspace-representation of $G$ over $\F$ if 
\begin{enumerate}
	\item $\dim S_v=d$ for all $v\in V$, and
	\item $S_v\cap\bigl(\sum_{u\not\sim v}S_u\bigr)=\{\bf{0}\}$, where the summation is over $u$ satisfying $uv\notin E$.
\end{enumerate}

With this, for a graph $G$ and a field $\F$, we have
	\[
		\subbound(G;\F)=\inf\biggl\{{n\over d}:\text{$G$ has an $(n,d)$-subspace-representation over $\F$}\biggr\}.
	\]
This formulation can be used to give a coordinate-free proof of Theorem~\ref{thm:tensorize}.

\section{Proofs of Theorems~\ref{thm:shannonbound} and~\ref{thm:tensorize} and Proposition~\ref{prop:cycles}}\label{sec:proof1}

We now set out to prove that $\Theta(G)\leq\subbound(G)$ and explore some basic properties.

Recalling the lexicographic product of graphs, the \emph{fractional chromatic number} of a graph $G$ is defined as
\[
\chi_f(G):=\inf_d{\chi(G\ltimes K_d)\over d}.
\]
In his original paper, Shannon~\cite{shannon1956zero} established $\Theta(G)\leq\chi_f(\ol{G})$.


In the same spirit, Lov\'asz showed that $\alpha(G)\leq\theta(G)\leq\chi_f(\ol{G})$. Similarly, Haemers established $\alpha(G)\leq\ham(G)\leq\chi(\ol{G})$; note that, in general, $\ham(G)\not\leq\chi_f(\ol{G})$, e.g.\ $\chi_f(\ol{C_5})=5/2$ whereas $\ham(C_5)\geq 3$.

\begin{theorem}\label{thm:indbound}
For any graph $G$,
\[
\alpha(G)\leq\subbound(G)\leq\chi_f(\ol{G}).
\]
\end{theorem}
\begin{proof}
Notice that for any graphs $G,H$, $\alpha(G\ltimes H)=\alpha(G)\cdot\alpha(H)$ and $\ol{G\ltimes H}=\ol{G}\ltimes\ol{H}$. Thus, as $\alpha(G)\leq\ham(G)\leq\chi(\ol{G})$, we find
\[
\alpha(G)=\inf_d{d\cdot\alpha(G)\over d}=\inf_d{\alpha(G\ltimes\ol{K_d})\over d}\leq\inf_d{\ham(G\ltimes\ol{K_d})\over d}=\subbound(G),
\]
and
\[
\subbound(G)=\inf_d{\ham(G\ltimes\ol{K_d})\over d}\leq\inf_d{\chi(\ol{G\ltimes\ol{K_d}})\over d}=\inf_d{\chi(\ol{G}\ltimes K_d)\over d}=\chi_f(\ol{G}).\qedhere
\]
\end{proof}

We now provide a proof of Theorem~\ref{thm:tensorize}. Before we do so, recall that $\theta(G\boxtimes H)=\theta(G)\cdot\theta(H)$; however, the same is not true of $\ham(G)$.

\begin{figure}[ht]
\begin{tikzpicture}		
\foreach \x in {0,1,2,3} {
	\foreach \y in {0,1,2,3} {
		\draw (\x,\y)--(\x+1,\y+1);}}
		
\foreach \x in {0,1,2,3} {
		\draw (\x,0)--(\x+1,4);
		\draw (\x,4)--(\x+1,0);
		\draw (0,\x)--(4,\x+1);
		\draw (4,\x)--(0,\x+1);}

\foreach \x in {1,2,3,4} {
	\foreach \y in {0,1,2,3} {
		\draw (\x,\y)--(\x-1,\y+1);}}
		
\foreach \x in {0,1,2,3,4} {
	\draw (\x,0)--(\x,4);
	\draw (\x,0) to [out=120,in=-120](\x,4);}
	
\foreach \y in {0,1,2,3,4} {
	\draw (0,\y)--(4,\y);
	\draw (0,\y) to [out=30,in=150](4,\y);}
	
\draw (0,0)to [out=60,in=210] (4,4);
\draw (0,4)to [out=-60,in=150] (4,0);

		
\foreach \x in {0,1} {
	\foreach \y in {0,1} {
		\node[vertex] at (\x,\y) {\textbf{7}};}}		

\foreach \x in {0,1} {
	\foreach \y in {3,4} {
		\node[vertex] at (\x,\y) {\textbf{1}};}}	

\foreach \x in {3,4} {
	\foreach \y in {3,4} {
		\node[vertex] at (\x,\y) {\textbf{3}};}}	
		
\foreach \x in {3,4} {
	\foreach \y in {0,1} {
		\node[vertex] at (\x,\y) {\textbf{8}};}}	
		
\foreach \x in {0,4} {
	\foreach \y in {2} {
		\node[vertex] at (\x,\y) {\textbf{4}};}}	
		
\foreach \x in {2} {
	\foreach \y in {0,4} {
		\node[vertex] at (\x,\y) {\textbf{2}};}}	
		
\foreach \x in {1,2} {
	\foreach \y in {2,3} {
		\node[vertex] at (\x,\y) {\textbf{5}};}}	
		
\node[vertex] at (2,1) {\textbf{6}};
\node[vertex] at (3,2) {\textbf{6}};
		
%
%
%
%
\end{tikzpicture}
\caption{A coloring of $C_5\boxtimes C_5$ using $8$ colors in which each color class is a clique.\label{fig:c5cover}}
\end{figure}
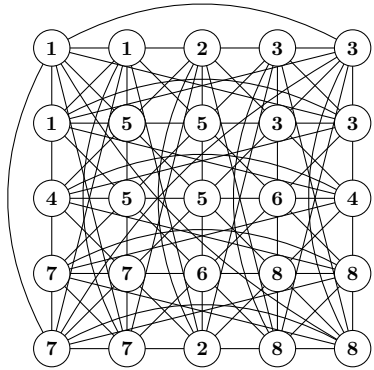

\begin{prop}\label{prop:hamnotens}
For any field $\F$, $\ham(C_5;\F)\geq 3$, yet $\ham(C_5\boxtimes C_5;\F)\leq 8$.
\end{prop}
\begin{proof}
As shown by Lov\'asz, $\Theta(C_5)=\sqrt{5}$. Thus, as $\ham$ is always an integer, $\ham(C_5;\F)\geq\lceil\sqrt{5}\rceil=3$.

On the other hand, it is not difficult to verify that $\chi(\ol{C_5\boxtimes C_5})\leq 8$; indeed, Figure~\ref{fig:c5cover} provides such a coloring. Therefore, $\ham(C_5\boxtimes C_5;\F)\leq 8$ for any field $\F$.
\end{proof}

\begin{proof}[Proof of Theorem~\ref{thm:tensorize}]
	\textit{Upper bound.} Let $M\in\fit_\F^{d_1}(G)$ and $N\in\fit_\F^{d_2}(H)$, and set $M^*=M\otimes N$ where $\otimes$ is the tensor/Kronecker product. We claim that $M^*\in\fit_\F^{d_1d_2}(G\boxtimes H)$. Indeed, the rows and columns of $M^*$ are indexed by $(V(G)\times[d_1])\times(V(H)\times[d_2])$, so we may in fact suppose they are indexed by $V(G\boxtimes H)\times[d_1d_2]$. Further, for any $(u,x),(v,y)\in V(G\boxtimes H)$, the $(u,x)(v,y)$ block of $M^*$ satisfies $M^*_{(u,x)(v,y)}=M_{uv}\otimes N_{xy}$. As such, $M^*_{(u,x)(u,x)}=I_{d_1}\otimes I_{d_2}=I_{d_1d_2}$. Further, if $(u,x)\not\sim(v,y)$ in $G\boxtimes H$, then either $uv\notin E(G)$ or $xy\notin E(H)$, so either $M_{uv}=M_{vu}=O_{d_1}$ or $N_{xy}=N_{yx}=O_{d_2}$. In either case, $M^*_{(u,x)(v,y)}=M^*_{(v,y)(u,x)}=O_{d_1d_2}$, so we have verified $M^*\in\fit_\F^{d_1d_2}(G\boxtimes H)$.

Finally, $\rank(M^*)=\rank(M)\cdot\rank(N)$, so
\[
\subbound(G\boxtimes H)\leq{\rank(M^*)\over d_1d_2}={\rank(M)\over d_1}\cdot{\rank(N)\over d_2}.
\]
Taking infimums establishes $\subbound(G\boxtimes H)\leq\subbound(G)\cdot\subbound(H)$.

\textit{Lower bound.} Let $M\in\fit_\F^d(G\boxtimes H)$ for some $d$. For $u,v\in V(G)$, let $[M]_{uv}$ denote the submatrix of $M$ consisting of all blocks of the form $M_{(u,x)(v,y)}$ for $x,y\in V(H)$. Certainly we can consider $M$ as a matrix with blocks $[M]_{uv}$, which is a $d_u\times d_v$ matrix for positive integers $d_u,d_v$. Additionally, for every $v\in V(G)$, the submatrix $[M]_{vv}$ can be considered a $d$\nobreakdash-representation of the graph $H$, where the $xy$ block is $([M]_{vv})_{xy}=M_{(v,x)(v,y)}$. Set $r=\min_{v\in V(G)}\rank([M]_{vv})$, so $\subbound(H)\leq r/d$. 

On the other hand, if $uv\notin E(G)$, then $[M]_{uv}=O_{d_u\times d_v}$ and $[M]_{vu}=O_{d_v\times d_u}$ as $(u,x)\not\sim(v,y)$ in $G\boxtimes H$ for every $x,y\in V(H)$. Thus, as $\rank([M]_{vv})\geq r$ for all $v\in V(G)$, $M$ is a rank-$r$-representation of $G$, so $\subbound(G)\leq\rank(M)/r$.

Putting the bounds on $\subbound(G)$ and $\subbound(H)$ together, we have
\[
\subbound(G)\cdot\subbound(H)\leq{\rank(M)\over r}\cdot{r\over d}={\rank(M)\over d},
\]
so taking infimums yields $\subbound(G)\cdot\subbound(H)\leq\subbound(G\boxtimes H)$.
\end{proof}

Putting together Theorems~\ref{thm:tensorize} and~\ref{thm:indbound}, we arrive at a proof of Theorem~\ref{thm:shannonbound}.
\begin{proof}[Proof of Theorem~\ref{thm:shannonbound}]
We have already noted that $\subbound(G)\leq\ham(G)$. On the other hand, by Theorems~\ref{thm:tensorize} and~\ref{thm:indbound},
\[
\Theta(G)=\sup_n\alpha(G^{\boxtimes n})^{1/n}\leq\sup_n\subbound(G^{\boxtimes n})^{1/n}=\sup_n\subbound(G)=\subbound(G).\qedhere
\]
\end{proof}

Theorem~\ref{thm:tensorize} has another nice corollary. Certainly $\Theta(G)\leq\ham(G)$, but one could additionally attain bounds on $\Theta(G)$ by using Haemers' bound on large powers of $G$, i.e.\ $\Theta(G)\leq\ham(G^{\boxtimes n})^{1/n}$. This could lead to improved bounds as in general $\ham(G^{\boxtimes 2})<\ham(G)^2$, e.g.\ $G=C_5$. It turns out that $\subbound(G)$ outperforms any bound attained in this fashion.
\begin{cor}\label{cor:beatslimit}
For any positive integer $n$ and graph $G$, $\subbound(G)\leq\ham(G^{\boxtimes n})^{1/n}$.
\end{cor}
\begin{proof}
By Theorem~\ref{thm:tensorize}, we calculate $\subbound(G)=\subbound(G^{\boxtimes n})^{1/n}\leq\ham(G^{\boxtimes n})^{1/n}$.
\end{proof}

To end this section, we show that the fractional Haemers bound cannot improve upon the known bounds for the Shannon capacity of odd cycles. We require the following observation about $(n,d)$\nobreakdash-representations of a graph.
\begin{prop}\label{prop:linind}
Let $G$ be a graph and $\{(A_v,B_v)\in(\F^{n\times d})^2:v\in V\}$ be an $(n,d)$-representation of~$G$. For $v\in V$, let $X_v$ denote the column space of $A_v$. If $S,T\subseteq V$ are disjoint sets of vertices where $S$ is an independent set and there are no edges between $S$ and $T$, then the subspaces $\sum_{v\in S}X_v$ and $\sum_{v\in T}X_v$ are linearly independent.
\end{prop}
\begin{proof}
Let $\{\mathbf{a}_{v,i}:i\in[d]\}$ be the columns of $A_v$ and $\{\mathbf{b}_{v,i}:i\in[d]\}$ be the columns of $B_v$. As $\{(A_v,B_v):v\in V\}$ is an $(n,d)$-representation of $G$, we know that 
\begin{enumerate}
\item $\langle\mathbf{a}_{v,i},\mathbf{b}_{v,i}\rangle=1$ for every $v\in V,i\in[d]$,
\item $\langle\mathbf{a}_{v,i},\mathbf{b}_{v,j}\rangle=0$ for every $v\in V$ and $i\neq j$, and
\item $\langle\mathbf{a}_{u,i},\mathbf{b}_{v,j}\rangle=0$ for every $i,j\in[d]$ whenever $uv\notin E$.
\end{enumerate}
Let $\cal{B}$ be a basis for $\sum_{v\in T}X_v$ and consider a linear combination
\[
\sum_{v\in S}\sum_{i=1}^d c_{v,i}\mathbf{a}_{v,i}+\sum_{\mathbf{x}\in\cal{B}}d_\mathbf{x}\mathbf{x}=0,
\]
where $c_{v,i},d_\mathbf{x}\in\F$ for every $v\in S,i\in[d],\mathbf{x}\in\cal{B}$. As $S$ is an independent set and there are no edges between $S$ and $T$, we find that for any $u\in S,j\in[d]$,
\[
0=\bigg\langle\sum_{v\in S}\sum_{i=1}^d c_{v,i}\mathbf{a}_{v,i}+\sum_{\mathbf{x}\in\cal{B}}d_\mathbf{x}\mathbf{x},\mathbf{b}_{u,j}\bigg\rangle=c_{u,j}.
\]
Therefore $c_{v,i}=0$ for every $v\in S,i\in[d]$, so we must have $\sum_{\mathbf{x}\in\cal{B}}d_\mathbf{x}\mathbf{x}=0$. As $\cal{B}$ is a basis, this implies $d_\mathbf{x}=0$ for every $\mathbf{x}\in\cal{B}$.

Thus, as $\sum_{v\in S}X_v=\Span_\F\{\mathbf{a}_{v,i}:v\in S,i\in[d]\}$ and $\sum_{v\in T}X_v=\Span_\F\cal{B}$, we have shown that $\sum_{v\in S}X_v$ and $\sum_{v\in T}X_v$ are linearly independent subspaces.
\end{proof}

\begin{proof}[Proof of Proposition~\ref{prop:cycles}]
We will show that $\subbound(C_{2k+1})=\chi_f(\ol{C_{2k+1}})=k+{1\over 2}$. It is well-known that $\chi_f(\ol{C_{2k+1}})=k+{1\over 2}$, so we will focus only on the lower bound.

Identify the vertices of $C_{2k+1}$ with $\Z_{2k+1}$ in the natural way and let $\{(A_i,B_i)\in(\F^{n\times d})^2:i\in\Z_{2k+1}\}$ be an $(n,d)$-representation of $C_{2k+1}$ for any $n,d$. Let $X_i$ denote the column space of $A_i$. As $A_i^TB_i=I_d$, we observe that $\dim(X_i)=d$.

We observe that $I=\{3,5,7,\dots,2k-1\}$ is an independent set in $C_{2k+1}$ and further that the edge $0\sim 1$ is not adjacent to any vertex in $I$. By iterating Proposition~\ref{prop:linind}, we find that 
\[ 
  \dim\biggl((X_0+X_1)+\sum_{i\in I}X_{i}\biggr)=\dim(X_0+X_1)+\dim\biggl(\sum_{i\in I}X_{i}\biggr)=\dim(X_0+X_1)+\sum_{i\in I}\dim(X_{i}),
\]
so
\begin{align*}
n &\geq \dim\biggl((X_0+X_1)+\sum_{i\in I}X_{i}\biggr)\\
&=\dim(X_0+X_1)+\sum_{i\in I}\dim(X_{i})\\
&=\dim(X_0)+\dim(X_1)-\dim(X_0\cap X_1)+\sum_{i\in I}\dim(X_{i})\\
&= (k+1)d-\dim(X_0\cap X_1).
\end{align*}
From this, we have $\dim(X_0\cap X_1)\geq (k+1)d-n$, and so by symmetry, for any $i\in\Z_{2k+1}$, $\dim(X_i\cap X_{i+1})\geq (k+1)d-n$. 

Because $1\not\sim 2k$ in $C_{2k+1}$, by Proposition~\ref{prop:linind} it follows that $X_1\cap X_{2k}=\{0\}$. 
Since we also have $(X_0\cap X_1)+(X_0\cap X_{2k})\leq X_0$, we conclude that
\begin{align*}
  d &=\dim(X_0)\geq\dim\bigl((X_0\cap X_1)+(X_0\cap X_{2k})\bigr)\\&=\dim(X_0\cap X_1)+\dim(X_0\cap X_{2k})\geq 2\bigl((k+1)d-n\bigr),
\end{align*}
which implies that
$
 {n\over d}\geq k+{1\over 2}.
$
Taking the infimum yields $\subbound(C_{2k+1})\geq k+{1\over 2}$.
\end{proof}

\section{Proof of Theorem~\ref{thm:itsbetter} and further separation}\label{sec:proof2}

In this section, we first give a proof of Theorem~\ref{thm:itsbetter}; namely, for every field $\F$ of nonzero characteristic, we need to find a graph $G=G(\F)$ for which $\subbound(G;\F)<\min\{\ham(G;\F'),\theta(G)\}$ for \emph{every} field $\F'$. After this, we provide further separation of $\subbound$ over fields of different characteristics. 

For the proof of Theorem~\ref{thm:itsbetter}, we need the following result. The first part of the following lemma was provided by Haemers in~\cite{haemers1978upper}; we provide a full proof for completeness.
\begin{lemma}\label{lem:ham}
For a prime $p$ and an integer $n$, let $J_{n}^p$ be the graph with vertex set ${[n]\choose p+1}$ where $X\sim Y$ in $J_{n}^p$ whenever $|X\cap Y|\not\equiv 0\pmod p$. 
\begin{enumerate}
\item If $(p+2)\mid n$ and $\F$ is a field of characteristic $p$, then
\[
\ham(J_{n}^p;\F)=\alpha(J_{n}^p)= n.
\]
\item For fixed $p$ and all large $n$,
\[
\theta(J_{n}^p)=\biggl({p\over(p+1)^2}+o(1)\biggr)n^2.
\]
\end{enumerate}
\end{lemma}
Before continuing with the proof, it is important to point out a typo in~\cite{haemers1978upper} in which it is stated that $\theta(J_n^2)={n(n-2)(2n-1)\over 3(3n-14)}$. 
The correct formula is $\theta(J_n^2)={n(n-2)(2n-11)\over 3(3n-14)}$, though for brevity's sake we prove only $\theta(J_n^2)\sim {2\over 9}n^2$.

\begin{proof}
\begin{enumerate}[leftmargin=*]
\item
Let $M\in\F^{[n]\times{[n]\choose p+1}}$ be the incidence matrix of all $(p+1)$-subsets of $[n]$, i.e.\ the matrix with entries $M_{i,X}=\mathbf{1}[i\in X]$. Certainly the matrix $M^TM$ fits $J_{n}^p$ over $\F$ as $\F$ has characteristic $p$ and $p+1\equiv 1\pmod p$. Thus, $\ham(J_{n}^p;\F)\leq\rank(M^TM)=\rank(M)\leq n$. On the other hand, as $(p+2)\mid n$, partition $[n]$ into sets $I_1,\dots,I_k$ where $|I_i|=p+2$ for all $i$. If $X,Y\subseteq I_i$ are sets of size $p+1$, then either $X=Y$ or $|X\cap Y|=p$. Thus, the collection ${I_1\choose p+1}\cup\dots\cup{I_k\choose p+1}$ is an independent set in $J_{n}^p$ and has size ${p+2\choose p+1}{n\over p+2}=n$.
\item We require the following fact which can be deduced quickly from~\cite{Schrijver_1981} (see specifically items (12), (13) and (27)): provided $n\geq 2(p+1)$,
\[
\begin{array}{ccccc}
\theta(J_n^p) & = & \max & 1+a_1+a_{p+1} \\
 & & \text{s.t.} & {(p+1-u)(n-p-u-1)-u\over (p+1)(n-p-1)}a_1+(-1)^u{{n-p-u-1\choose p+1-u}\over{n-p-1\choose p+1}}a_{p+1}\geq -1, & \text{for $u\in\{0,\dots,p+1\}$}.
\end{array}
\]

\noindent For large $n$, the preceding inequality can be written as
\begin{align}
	\label{eq:initial}  \bigl(\tfrac{p+1-u}{p+1}+o(1)\bigr)a_1+\bigl(\tfrac{(p+1)!}{(p+1-u)!}+o(1)\bigr)\tfrac{(-1)^u}{n^u}a_{p+1}&\geq -1\qquad\text{for }u\in\{0,\dotsc,p\}\\
	\label{eq:final}   -\bigl(1+o(1)\bigr)\tfrac{1}{n}a_1+\bigl((p+1)!+o(1)\bigr)\tfrac{(-1)^{p+1}}{n^{p+1}}a_{p+1}&\geq -1\qquad\text{for }u=p+1,
\end{align}
where $o(1)\to 0$ as $n\to\infty$.

\noindent Set $a_1=n+o(n)$ where the precise value of the $o(n)$ term is chosen so that \eqref{eq:final} is satisfied. Then
set $a_{p+1}=\frac{p}{(p+1)^2}n^2+o(n^2)$ where the value of $o(n^2)$ is chosen so the inequality \eqref{eq:initial} with $u=1$ is satisfied.
The remaining inequalities are then satisfied as well. Indeed, for even $u$, both terms on the left side of \eqref{eq:initial} are positive,
whereas for odd $u\geq 3$ the second term is $o(1)$ for $n\to\infty$. Hence, $\theta(J_n^p)\geq \bigl({p\over(p+1)^2}+o(1)\bigr)n^2$.\vspace{5pt}

\noindent On the other hand, when $p=2$, \eqref{eq:final} immediately implies that $a_1\leq n+o(n)$. 
For $p>2$, we also find that $a_1\leq n+o(n)$ by putting together \eqref{eq:final} and the $u=1$ case of \eqref{eq:initial}.

\noindent In either case, the $u=1$ case of \eqref{eq:initial} implies that
\[
	a_{p+1}\leq \bigl(1+o(1)\bigr)\tfrac{1+\bigl(\tfrac{p}{p+1}+o(1)\bigr)a_1}{p+1}n\leq\bigl(\tfrac{p}{(p+1)^2}+o(1)\bigr)n^2.
\]
Hence, $\theta(J_n^p)\leq 1+a_1+a_{p+1}\leq\bigl({p\over (p+1)^2}+o(1)\bigr)n^2$.\qedhere
\end{enumerate}
\end{proof}

\begin{proof}[Proof of Theorem~\ref{thm:itsbetter}]
Let $\F$ be a field of characteristic $p$ and set $G=J_{n}^p\boxtimes C_5^{\boxtimes 3}$ where $(p+2)\mid n$ and $8\nmid n$. By Proposition~\ref{prop:cycles} and Lemma~\ref{lem:ham}, we calculate
\[
\subbound(G;\F)=\subbound(J_{n}^p;\F)\cdot\subbound(C_5;\F)^3={125\over 8}n.
\]
For \emph{any} field $\F'$, as $n=\alpha(J_{n}^p)\leq\subbound(J_{n}^p;\F')$ and $\subbound(C_5;\F')={5\over 2}$, we have 
\[
\ham(G;\F')\geq\subbound(G;\F')=\subbound(J_n^p;\F')\cdot\subbound(C_5;\F')^3\geq{125\over 8}n.
\]
 However, as $8\nmid n$ and $\ham$ is always an integer, we have $\ham(G;\F')\geq\lceil{125\over 8}n\rceil>\subbound(G;\F)$.

Further, by Lemma~\ref{lem:ham} and the fact that $\theta(C_5)=\sqrt{5}$, we have
\[
\theta(G)=\theta(J_{n}^p)\cdot\theta(C_5)^3= 5^{3/2}\biggl({p\over(p+1)^2}+o(1)\biggr)n^2.
\]
Thus, for sufficiently large $n$ with $(p+2)\mid n$ and $8\nmid n$, 
\[
\subbound(G;\F)<\min\{\ham(G;\F'),\theta(G)\},
\]
for every field $\F'$.
\end{proof}

We next show that the choice of field matters when evaluating $\subbound$. In particular, for any field $\F$ of nonzero characteristic, we will show that there is an explicit graph $G=G(\F)$ for which $\subbound(G;\F)<\subbound(G;\F')$ for any field $\F'$ with $\cha(\F')\neq\cha(\F)$.

First, we define a ``universal graph'' for $\subbound(G;\F)$. For a field $\F$ and positive integers $d\leq n$, define the graph $\uni_\F(n,d)$ as follows: $V(\uni_\F(n,d))=\big\{(A,B)\in(\F^{n\times d})^2:A^TB=I_d\big\}$ where $(A,B)\not\sim(C,D)$ in $\uni_\F(n,d)$ if and only if $A^TD=C^TB=O_d$. We require the following two facts.

\begin{obs}\label{obs:homo}
For graphs $G$ and $H$, if there is a graph homomorphism from $\ol{G}$ to $\ol{H}$, then $\subbound(G)\leq\subbound(H)$.
\end{obs}

\begin{obs}\label{obs:uni}
A graph $G$ has an $(n,d)$-representation over $\F$ if and only if there is a graph homomorphism from $\ol{G}$ to $\ol{\uni_\F(n,d)}$. In particular, if $\ham(G;\F)=n$, then there is a graph homomorphism from $\ol{G}$ to $\ol{\uni_\F(n,1)}$.
\end{obs}

Further, we can essentially pin down the Shannon capacity of $\uni_\F(n,d)$.
\begin{prop}
For positive integers $d\leq n$ and a field $\F$,
\[
\bigg\lfloor{n\over d}\bigg\rfloor=\alpha(\uni_\F(n,d))\leq\Theta(\uni_\F(n,d))\leq\subbound(\uni_\F(n,d);\F)\leq{n\over d}.
\]
\end{prop}
\begin{proof}
By definition, the vertices of $\uni_\F(n,d)$ are their own $(n,d)$-representation over $\F$; therefore $\subbound(\uni_\F(n,d);\F)\leq{n\over d}$.

On the other hand, certainly $\big\lfloor{n\over d}\big\rfloor\leq\alpha(\uni_\F(n,d))$ by considering the independent set made of the pairs $\{(A_i,A_i):1\leq i\leq\lfloor n/d\rfloor\}$ where $A_i=\big[\mathbf{e}_{(i-1)d+1}|\cdots|\mathbf{e}_{id}\big]$.

The full claim follows from the fact that $\alpha(G)\leq\Theta(G)\leq\subbound(G;\F)$.
\end{proof}

We also require the following formulation of $\ham(G;\F)$ given by Alon in \cite{alon1998shannon}.
\begin{obs}
For a graph $G$ and a field $\F$, consider assigning to each $v\in V$ both a polynomial and vector $(P_v,\mathbf{x}_v)\in\F[x_1,\dots,x_n]\times\F^n$. Such a collection of pairs is said to represent~$G$ if $P_v(\mathbf{x}_v)\neq\nobreak 0$ for every $v\in V$ and $P_u(\mathbf{x}_v)=P_v(\mathbf{x}_u)=0$ whenever $uv\notin E$. If $\{(P_v,\mathbf{x}_v):v\in V\}$ represents~$G$, then $\ham(G;\F)\leq\dim\Span_\F\{P_v:v\in V\}$. 
\end{obs}

A proof of the above observation follows from considering the matrix $M=(m_{uv})\in\F^{V\times V}$ with entries $m_{uv}=(P_u(\mathbf{x}_u))^{-1}P_u(\mathbf{x}_v)$.

We will also require a graph used by Alon in \cite{alon1998shannon}. For primes $p,q$ and a positive integer $n$, define the graph $B_n^{p,q}$ as follows: $V(B_n^{p,q})={[n]\choose pq-1}$ and $X\sim Y$ in $B_n^{p,q}$ if and only if $|X\cap Y|\equiv -1\pmod p$. We will make use of $B_n^{p,q}$ both when $p=q$ and when $p$ and $q$ are distinct.

The majority of the following was proved in \cite{alon1998shannon}, but we provide a full proof for completeness.
\begin{prop}\label{prop:alongraph}
Let $p$ and $q$ be (not necessarily distinct) primes.
\begin{enumerate}
\item If $\F$ is a field of characteristic $p$, then
\[
\ham(B_n^{p,q};\F)\leq\sum_{i=0}^{p-1}{n\choose i}.
\]

\item If $q\neq p$ and $\F'$ is a field of characteristic $q$, then
\[
\ham(\ol{B_n^{p,q}};\F')\leq \sum_{i=0}^{q-1}{n\choose i}.
\]

\item If $\F''$ is any field with either $\cha(\F'')=0$ or $\cha(\F'')>p$, then
\[
\ham(\ol{B_n^{p,p}};\F'')\leq \sum_{i=0}^{p-1}{n\choose i}.
\]
\end{enumerate}
\end{prop}
\begin{proof}
\begin{enumerate}[leftmargin=*]
\item For $X\in{[n]\choose pq-1}$, define the polynomial $P_X\in\F[\mathbf{x}]$ in $n$ variables by 
\[
P_X(\mathbf{x})=\prod_{i=0}^{p-2}\bigl(\langle \mathbf{1}_X, \mathbf{x}\rangle-i\bigr),
\]
where $\mathbf{1}_X$ is the indicator vector of the set $X$. Notice that $|X|=pq-1\equiv -1\pmod p$, so $P_X(\mathbf{1}_X)\neq 0$. Additionally, if $Y\in{[n]\choose pq-1}$ has $|X\cap Y|\not\equiv -1\pmod p$, then $P_X(\mathbf{1}_Y)=0$ over $\F$. Reducing $P_X$ to the multilinear polynomial $\hat{P}_X$ by repeatedly applying the identity $x^2=x$, we notice that $\hat{P}_X$ also has the properties stated above. Thus, $\big\{(\hat{P}_X,\mathbf{1}_X):X\in{[n]\choose pq-1}\big\}$ represents $B_n^{p,q}$. Each $\hat{P}_X$ is a multilinear polynomial in $n$ variables of degree at most $p-1$, so
\[
\ham(B_n^{p,q};\F)\leq\dim\Span_\F\bigg\{\hat{P}_X:X\in{[n]\choose pq-1}\bigg\}\leq\sum_{i=0}^{p-1}{n\choose i}.
\]
\item For $X\in{[n]\choose pq-1}$, define the polynomial $Q_X\in\F'[\mathbf{x}]$ in $n$ variables by 
\[Q_X(\mathbf{x})=\prod_{i=0}^{q-2}\bigl(\langle\mathbf{1}_X,\mathbf{x}\rangle-i\bigr).\] As $|X|=pq-1\equiv -1\pmod q$, we have $Q_X(\mathbf{1}_X)\neq 0$. Additionally, if $Y\in{[n]\choose pq-1}\setminus\{X\}$ has $|X\cap Y|\equiv -1\pmod p$, then as $|X\cap Y|<pq-1$ and $q\neq p$, we must have $|X\cap Y|\not\equiv -1\pmod q$. Therefore, $Q_X(\mathbf{1}_Y)=0$ over $\F'$. Again, reducing $Q_X$ to the multilinear polynomial $\hat{Q}_X$, we find that $\big\{(\hat{Q}_X,\mathbf{1}_X):X\in{[n]\choose pq-1}\big\}$ represents $\ol{B_n^{p,q}}$. As each $\hat{Q}_X$ is a multilinear polynomial in $n$ variables of degree at most $q-1$,
\[
\ham(\ol{B_n^{p,q}};\F')\leq\dim\Span_{\F'}\bigg\{\hat{Q}_X:X\in{[n]\choose pq-1}\bigg\}\leq\sum_{i=0}^{q-1}{n\choose i}.
\]
\item For $X\in{[n]\choose p^2-1}$, define the polynomial $R_X\in\F''[\mathbf{x}]$ in $n$ variables by 
\[
  R_X(\mathbf{x})=\prod_{i=1}^{p-1}\bigl(\langle\mathbf{1}_X,\mathbf{x}\rangle-(ip-1)\bigr).
\]
We notice that $R_X(\mathbf{1}_X)=\prod_{i=1}^{p-1}\bigl((p^2-1)-(ip-1)\bigr)=p^{p-1}(p-1)!$. Thus, as $\cha(\F'')=\nobreak 0$ or $\cha(\F'')>p$, we have $R_X(\mathbf{1}_X)\neq 0$ over $\F''$. Furthermore, whenever $Y\in{[n]\choose p^2-1}\setminus\{X\}$ and $|X\cap Y|\equiv\nobreak -1\pmod p$, we have $R_X(\mathbf{1}_Y)=0$. Finally, reducing $R_X$ to the multilinear polynomial $\hat{R}_X$, we know that $\big\{(\hat{R}_X,\mathbf{1}_X):X\in{[n]\choose p^2-1}\big\}$ represents $\ol{B_n^{p,p}}$, so
\[
\ham(\ol{B_n^{p,p}};\F'')\leq\dim\Span_{\F''}\bigg\{\hat{R}_X:X\in{[n]\choose p^2-1}\bigg\}\leq\sum_{i=0}^{p-1}{n\choose i}.\qedhere
\]
\end{enumerate}
\end{proof}

\begin{lemma}\label{lem:fracalongraph}
For distinct primes $p,q$ and a number $\epsilon>0$, there is an integer $n_{p,q}=n_{p,q}(\epsilon)$ so that if $\F$ is a field of characteristic $p$ and $\F'$ is a field of characteristic $q$, then whenever $n\geq n_{p,q}$,
\[
\ham(B_n^{p,q};\F)<\epsilon\cdot\ham_f(B_n^{p,q};\F').
\]
Further, there is an integer $n_{p,p}=n_{p,p}(\epsilon)$ so that if $\F''$ is any field with either $\cha(\F'')=0$ or $\cha(\F'')>p$, then whenever $n\geq n_{p,p}$,
\[
\ham(B_n^{p,p};\F)<\epsilon\cdot\ham_f(B_n^{p,p};\F'').
\]
\end{lemma}
\begin{proof}
Set $B=B_n^{p,q}$ and suppose that $\ham(B;\F)\geq\epsilon\cdot\ham_f(B;\F')$. For any graph $G$, $\alpha(G\boxtimes\ol{G})\geq|V(G)|$, so by Proposition~\ref{prop:alongraph},
\[
{n\choose pq-1}\leq\alpha(B\boxtimes\ol{B})\leq\ham_f(B\boxtimes\ol{B};\F')=\ham_f(B;\F')\cdot\ham_f(\ol{B};\F')\leq{1\over\epsilon}\sum_{i=0}^{p-1}{n\choose i}\sum_{i=0}^{q-1}{n\choose i}.
\]
The left-hand side is a polynomial of degree $pq-1$ whereas the right-hand side is a polynomial of degree $p+q-2<pq-1$; a contradiction for all sufficiently large $n$ compared to $p,q$.

Similarly, set $B=B_n^{p,p}$ and suppose that $\ham(B;\F)\geq\epsilon\cdot\ham_f(B;\F'')$, then by Proposition~\ref{prop:alongraph},
\[
{n\choose p^2-1}\leq\alpha(B\boxtimes\ol{B})\leq\ham_f(B\boxtimes\ol{B};\F'')=\ham_f(B;\F'')\cdot\ham_f(\ol{B};\F'')\leq{1\over\epsilon}\biggl(\sum_{i=0}^{p-1}{n\choose i}\biggr)^2;
\]
another contradiction for all sufficiently large $n$ compared to $p$.
\end{proof}

\begin{theorem}\label{thm:separate}
For any field $\F$ of nonzero characteristic and $\epsilon>0$, there exists an explicit graph $G=G(\F,\epsilon)$ so that if $\F'$ is any field with $\cha(\F')\neq\cha(\F)$, then
\[
\subbound(G;\F)<\epsilon\cdot\subbound(G;\F').
\]
\end{theorem}
\begin{proof}
Suppose that $\cha(\F)=p$ for some prime $p$ and set $n=\max\{n_{p,q}:q\leq p,\text{ $q$ prime}\}$ where $n_{p,q}=n_{p,q}(\epsilon)$ is as in Lemma~\ref{lem:fracalongraph}. For this $n$, set $N_{p,q}=\ham(B_n^{p,q};\F)$ and set $G_{p,q}=\uni_\F(N_{p,q},1)$. We know that
\[
N_{p,q}=\alpha(G_{p,q})=\ham(G_{p,q};\F)=\subbound(G_{p,q};\F).
\]
Define
\[
G=\prod_{\substack{q\leq p:\\ \text{$q$ prime}}}G_{p,q},
\]
where the product is the strong product. Notice that $n_{p,q}$ depends only on $p,q,\epsilon$, so $G$ depends only on $\epsilon$ and on the field $\F$.

Now, if $\F'$ is any field with $\cha(\F')\neq p$, then by Lemma~\ref{lem:fracalongraph} and the choice of $n$, there is some prime $q^*\leq p$, for which $N_{p,q^*}=\ham(B_n^{p,q^*};\F)<\epsilon\cdot\subbound(B_n^{p,q^*};\F')\leq\epsilon\cdot\subbound(G_{p,q^*};\F')$, where the last inequality follows from the fact that $\ham(B_n^{p,q^*};\F)=N_{p,q^*}$, so there is a graph homomorphism from $\ol{B_n^{p,q^*}}$ to $\ol{G_{p,q^*}}$ (see Observations~\ref{obs:homo} and~\ref{obs:uni}).

Additionally, for all other $q\leq p$, we have $\subbound(G_{p,q};\F')\geq\alpha(G_{p,q})=N_{p,q}$. Therefore,
\begin{align*}
\subbound(G;\F) &= \prod_{\substack{q\leq p:\\ \text{$q$ prime}}}\subbound(G_{p,q};\F)= \prod_{\substack{q\leq p:\\ \text{$q$ prime}}} N_{p,q}\\
&< \epsilon\cdot \subbound(G_{p,q^*};\F')\prod_{\substack{q\leq p,\ q\neq q^*:\\ \text{$q$ prime}}}N_{p,q}\\
&\leq \epsilon\cdot\prod_{\substack{q\leq p:\\ \text{$q$ prime}}}\subbound(G_{p,q};\F')=\epsilon\cdot\subbound(G;\F').\qedhere
\end{align*}
\end{proof}

\section{Fractionalizing Lov\'asz's theta function}\label{sec:lovasz}

One could attempt to fractionalize Lov\'asz's theta function in ways similar to how we fractionalized Haemers' bound. In this section, we provide two attempts and show that neither yields any improvements.

Recall that for a graph $G$, a collection of unit vectors $\{\mathbf{x}_v\in\R^n:v\in V\}$ is said to be an \emph{orthonormal representation of $G$} if $\langle \mathbf{x}_u,\mathbf{x}_v\rangle=0$ whenever $uv\notin E$. A \emph{handle} is simply a unit vector $\mathbf{h}$. The theta function of $G$ is defined to be
\[
\theta(G)=\min\max_{v\in V}{1\over\langle\mathbf{x}_v,\mathbf{h}\rangle^2}
\]
where the minimum is taken over all $\{\mathbf{x}_v:v\in V\}$, which are orthonormal representations of $G$, and all handles $\mathbf{h}$.
%

Recall also that $\theta(G)=\max\sum_{v\in V}\langle\bar{\mathbf{x}}_v,\bar{\mathbf{h}}\rangle^2$ where the maximum is taken over all $\{\bar{\mathbf{x}}_v:v\in V\}$, which are orthonormal representations of $\ol{G}$, and all handles $\bar{\mathbf{h}}$. This ``dual form'' of the theta function will be essential below.

A first attempt at fractionalizing the theta function is to define
\[
\theta_f(G):=\inf_d{\theta(G\ltimes\ol{K_d})\over d}.
\]
We recover $\theta(G)$ when $d=1$, so certainly $\theta_f(G)\leq\theta(G)$. Unfortunately, it turns out that $\theta_f$ is equal to~$\theta$.

\begin{theorem}
For any graph $G$, $\theta_f(G)=\theta(G)$.
\end{theorem}
\begin{proof}
As noted above, $\theta_f(G)\leq\theta(G)$. On the other hand, let $\{\bar{\mathbf{x}}_v:v\in V\}$ be an orthonormal representation of $\ol{G}$ and $\bar{\mathbf{h}}$ be a handle for which $\theta(G)=\sum_{v\in V}\langle\bar{\mathbf{x}}_v,\bar{\mathbf{h}}\rangle^2$. Let $d$ be any positive integer and for $i\in[d]$ and $v\in V$ define $\bar{\mathbf{x}}_{(v,i)}=\bar{\mathbf{x}}_v$. Certainly $\{\bar{\mathbf{x}}_{(v,i)}:v\in V,i\in[d]\}$ is an orthonormal representation for $\ol{G}\ltimes K_d=\ol{G\ltimes\ol{K_d}}$, so
\[
\theta(G\ltimes\ol{K_d})\geq\sum_{v\in V,i\in[d]}\langle\bar{\mathbf{x}}_{(v,i)},\bar{\mathbf{h}}\rangle^2=d\cdot\sum_{v\in V}\langle\bar{\mathbf{x}}_v,\bar{\mathbf{h}}\rangle^2=d\cdot\theta(G).
\]
Therefore,
\[
\theta_f(G)=\inf_d{\theta(G\ltimes\ol{K_d})\over d}\geq\inf_d{d\cdot\theta(G)\over d}=\theta(G).\qedhere
\]
\end{proof}
As a brief note, the above provides a quick proof that $\theta(G\ltimes\ol{K_d})=d\cdot\theta(G)$ for every positive integer $d$.

A second attempt may proceed by replacing vectors by matrices. In particular, for some positive integer $n$, we say that a collection of matrices $\{M_v\in\R^{n\times d_v}:v\in V\}$, where $d_v$ is some positive integer assigned to $v$, is a \emph{matrix representation} of $G$ if $M_v^TM_v=I_d$ for every $v\in V$ and $M_u^TM_v=O_d$ whenever $uv\notin E$. A \emph{$k$-handle} is a matrix $H\in\R^{n\times k}$ with $\Vert\mathbf{h}_i\Vert=1$ for all $i\in[k]$ where $\mathbf{h}_i$ is the $i$th column of $H$.

For a positive integer $k$, we define
\[
\theta_f(G):=\inf_{k\in\Z^+}\quad\inf_{\substack{\{M_v\in\R^{n\times d_v}:v\in V\}:\\ \text{a matrix representation of $G$}}}\quad \inf_{\substack{H\in\R^{n\times k}:\\ \text{a $k$-handle}}}\quad\max_{v\in V}\quad{k\over\tr(M_v^THH^TM_v)}.
\]
Unlike in the definition of $\theta(G)$, it is not clear, a priori, that the infimums can be replaced by minimums. 

We recover $\theta(G)$ when $k=d_v=1$ for all $v\in V$, so $\theta_f(G)\leq\theta(G)$. Sadly, yet again, it turns out that $\theta_f$ is equal to $\theta$.
\begin{theorem}
For any graph $G$, $\theta_f(G)=\theta(G)$.
\end{theorem}
\begin{proof}
We have already noted that $\theta_f(G)\leq\theta(G)$, so we need only establish the opposite inequality. The proof hinges on the following lemma.
\begin{lemma}
Let $\{M_v\in\R^{n\times d_v}:v\in V\}$ be a matrix representation of $G$ and $\{\bar{M}_v\in\R^{\bar{n}\times\bar{d}_v}:v\in V\}$ be a matrix representation of $\ol{G}$. For any $k$-handle $H\in\R^{n\times k}$ and any $\bar{k}$-handle $\bar{H}\in\R^{\bar{n}\times\bar{k}}$,
\[
k\bar{k}\geq\sum_{v\in V}\tr(M_v^THH^TM_v)\tr(\bar{M}_v^T\bar{H}\bar{H}^T\bar{M}_v).
\]
\end{lemma}
\begin{proof}[Proof of Lemma]
Let $\{\mathbf{h}_i:i\in[k]\}$ be the columns of $H$, $\{\bar{\mathbf{h}}_i:i\in[\bar{k}]\}$ be the columns of $\bar{H}$, $\{\mathbf{m}_{v,i}:i\in[d_v]\}$ be the columns of $M_v$ and $\{\bar{\mathbf{m}}_{v,i}:i\in[\bar{d}_v]\}$ be the columns of $\bar{M}_v$. We begin by noticing that for any $u,v\in V$ and $s\in[d_u],s'\in[d_v],t\in[\bar{d}_u],t'\in[\bar{d}_v]$, we have
\[
\langle \mathbf{m}_{u,s}\otimes\bar{\mathbf{m}}_{u,t},\mathbf{m}_{v,s'}\otimes\bar{\mathbf{m}}_{v,t'}\rangle=\begin{cases}
1 & \text{if $(u,s,t)=(v,s',t')$,}\\
0 & \text{otherwise}.
\end{cases}
\]
Firstly, if $u=v$, the claim follows from the fact that $\{\mathbf{m}_{v,i}:i\in[d_v]\}$ and $\{\bar{\mathbf{m}}_{v,i}:i\in[\bar{d}_v]\}$ are orthonormal. If $u\neq v$, then either $uv\notin E(G)$ or $uv\notin E(\bar{G})$, so  either $\langle \mathbf{m}_{u,s},\mathbf{m}_{v,s'}\rangle=0$ or $\langle\bar{\mathbf{m}}_{u,t},\bar{\mathbf{m}}_{v,t'}\rangle=0$. 

Therefore, the vectors $\{\mathbf{m}_{v,s}\otimes\bar{\mathbf{m}}_{v,t}:v\in V,s\in[d_v],t\in[\bar{d}_v]\}$ are orthonormal. From this, we calculate
{\allowdisplaybreaks
\begin{align*}
k\bar{k} &= \biggl(\sum_{i=1}^k \langle \mathbf{h}_i,\mathbf{h}_i\rangle^2\biggr)\biggl(\sum_{j=1}^{\bar{k}}\langle\bar{\mathbf{h}}_j,\bar{\mathbf{h}}_j\rangle^2\biggr)\\
&= \sum_{i=1}^k\sum_{j=1}^{\bar{k}} \langle \mathbf{h}_i,\mathbf{h}_i\rangle^2\langle\bar{\mathbf{h}}_j,\bar{\mathbf{h}}_j\rangle^2\\
&= \sum_{i=1}^k\sum_{j=1}^{\bar{k}}\langle \mathbf{h}_i\otimes\bar{\mathbf{h}}_j,\mathbf{h}_i\otimes\bar{\mathbf{h}}_j\rangle^2\\
&\geq \sum_{i=1}^k\sum_{j=1}^{\bar{k}}\sum_{v\in V}\sum_{s=1}^{d_v}\sum_{t=1}^{\bar{d}_v}\langle \mathbf{h}_i\otimes\bar{\mathbf{h}}_j,\mathbf{m}_{v,s}\otimes\bar{\mathbf{m}}_{v,t}\rangle^2\\
&=\sum_{i=1}^k\sum_{j=1}^{\bar{k}}\sum_{v\in V}\sum_{s=1}^{d_v}\sum_{t=1}^{\bar{d}_v}(\mathbf{m}_{v,s}\otimes\bar{\mathbf{m}}_{v,t})^T(\mathbf{h}_i\otimes\bar{\mathbf{h}}_j)(\mathbf{h}_i\otimes\bar{\mathbf{h}}_j)^T(\mathbf{m}_{v,s}\otimes\bar{\mathbf{m}}_{v,t})\\
&= \sum_{i=1}^k\sum_{j=1}^{\bar{k}}\sum_{v\in V}\sum_{s=1}^{d_v}\sum_{t=1}^{\bar{d}_v}\tr\bigl((\mathbf{h}_i\otimes\bar{\mathbf{h}}_j)(\mathbf{h}_i\otimes\bar{\mathbf{h}}_j)^T(\mathbf{m}_{v,s}\otimes\bar{\mathbf{m}}_{v,t})(\mathbf{m}_{v,s}\otimes\bar{\mathbf{m}}_{v,t})^T\bigr)\\
&=\sum_{v\in V}\tr\biggl(\biggl(\sum_{i=1}^k\sum_{j=1}^{\bar{k}}(\mathbf{h}_i\otimes\bar{\mathbf{h}}_j)(\mathbf{h}_i\otimes\bar{\mathbf{h}}_j)^T\biggr)\biggl(\sum_{s=1}^{d_v}\sum_{t=1}^{\bar{d}_v}(\mathbf{m}_{v,s}\otimes\bar{\mathbf{m}}_{v,t})(\mathbf{m}_{v,s}\otimes\bar{\mathbf{m}}_{v,t})^T\biggr)\biggr)\\
&=\sum_{v\in V}\tr\bigl((H\otimes\bar{H})(H\otimes\bar{H})^T(M_v\otimes\bar{M}_v)(M_v\otimes\bar{M}_v)^T\bigr)\\
&=\sum_{v\in V}\tr\bigl((M_v\otimes\bar{M}_v)^T(H\otimes\bar{H})(H\otimes\bar{H})^T(M_v\otimes\bar{M}_v)\bigr)\\
&=\sum_{v\in V}\tr\bigl((M_v^THH^TM_v)\otimes(\bar{M}_v^T\bar{H}\bar{H}^T\bar{M}_v)\bigr)\\
&=\sum_{v\in V}\tr(M_v^THH^TM_v)\tr(\bar{M}_v^T\bar{H}\bar{H}^T\bar{M}_v).\qedhere
\end{align*}}
\end{proof}

Using the above lemma, the theorem follows quickly. Let $\{\bar{\mathbf{x}}_v:v\in V\}$ be an orthonormal representation of $\ol{G}$ and $\bar{\mathbf{h}}$ be a handle with $\theta(G)=\sum_{v\in V}\langle\bar{\mathbf{x}}_v,\bar{\mathbf{h}}\rangle^2$. Also, for any $\epsilon>0$, let $\{M_v:v\in V\}$ be a matrix representation of $G$ and $H$ be a $k$-handle with $\theta_f(G)+\epsilon\geq\max_{v\in V}{k\over\tr(M_v^THH^TM_v)}$. By the lemma, 
\[
k\cdot 1\geq\sum_{v\in V}\tr(M_v^THH^TM_v)\tr(\bar{\mathbf{x}}_v^T\bar{\mathbf{h}}\bar{\mathbf{h}}^T\bar{\mathbf{x}}_v)\geq{k\over\theta_f(G)+\epsilon}\sum_{v\in V}\langle\bar{\mathbf{x}}_v,\bar{\mathbf{h}}\rangle^2=k\cdot{\theta(G)\over\theta_f(G)+\epsilon}.
\]
Thus, $\theta_f(G)+\epsilon\geq\theta(G)$ for every $\epsilon>0$, so $\theta_f(G)\geq\theta(G)$.
\end{proof}

Although it turns out that the matrix formulation of $\theta_f(G)$ does not provide any improvements on $\theta(G)$, it could still be useful in providing bounds on $\theta(G)$ for large graphs or establishing general theorems. For example, a theorem of Lov\'asz in \cite{lovasz1979shannon} states that if $G$ has an orthonormal representation in $\R^N$, then $\theta(G)\leq N$. While this is not difficult to prove directly from the definition of $\theta(G)$, it does require some creativity; however, it follows essentially by definition for $\theta_f(G)$. In fact, we can quickly show something stronger. For a graph $G$, a collection of subspaces $\{S_v\leq\R^N:v\in V\}$ is said to be a $d$\nobreakdash-dimensional representation over $\R^N$ if $\dim(S_v)=d$ for every $v\in V$ and $S_u\perp S_v$ whenever $uv\notin E$. Of course, a $1$-dimensional representation is equivalent to an orthonormal representation.
\begin{prop}
If $G$ has a $d$-dimensional representation over $\R^N$, then $\theta(G)\leq N/d$.
\end{prop}
\begin{proof}
Let $\{S_v\leq\R^N:v\in V\}$ be a $d$-dimensional representation of $G$ over $\R^N$. For each $v\in V$, let $M_v\in\R^{N\times d}$ be a matrix whose columns form an orthonormal basis for $S_v$; thus, $M_v^TM_v=I_d$. Further, as $S_u\perp S_v$ whenever $uv\notin E$, we have $M_u^TM_v=O_d$ whenever $uv\notin E$, so $\{M_v:v\in V\}$ is a matrix representation of $G$. Let $H=I_N$, which is an $N$-handle, so
\[
\theta(G)=\theta_f(G)\leq\max_{v\in V}{N\over\tr(M_v^THH^TM_v)}=\max_{v\in V}{N\over\tr(I_d)}={N\over d}.\qedhere
\]
\end{proof}

\section{Conclusion}\label{sec:conclusion}

We conclude with a list of open questions related to our study of $\subbound(G)$.
\begin{itemize}[leftmargin=*,itemsep=4pt]
\item For a graph $G$ and a field $\F$, is $\subbound(G;\F)$ attained? That is to ask: is the infimum really a minimum? Beyond this, is $\subbound(G;\F)$ computable? 

\item Theorem~\ref{thm:itsbetter} shows that for any field $\F$ of nonzero characteristic, there is a graph $G=G(\F)$ with $\subbound(G;\F)<\min\{\ham(G;\F'),\theta(G)\}$ for every field $\F'$. While this $G$ satisfied $\subbound(G;\F)<\epsilon\cdot\theta(G)$, we only verified that $\subbound(G;\F)\leq\ham(G;\F)-{1\over 8}$. It would nice to construct a graph $G$ with $\subbound(G;\F)<\epsilon\cdot\min\{\ham(G;\F'),\theta(G)\}$ for every field $\F'$. We believe that such a graph does indeed exist.

\item Theorem~\ref{thm:separate} shows that for any field $\F$ of nonzero characteristic, there is a graph $G=G(\F)$ with $\subbound(G;\F)<\subbound(G;\F')$ for every field $\F'$ with $\cha(\F')\neq\cha(\F)$. 
We were unable to prove a similar separation for fields of equal characteristic. Namely,
given a \emph{finite} field $\F$, is there a graph $G=G(\F)$ with $\subbound(G;\F)<\subbound(G;\F')$ for every field $\F'$ that is not an extension of $\F$? We suspect that the graph $\uni_\F(n,d)$ provides such an example for appropriately chosen~$n,d$.

\item Are there graphs for which $\theta(G)<\ham(G)$, yet $\subbound(G)<\theta(G)$? 
	While this paper was in submission, this question was answered affirmatively by Hu, Tamo and Shayevitz in~\cite{hts18} using their parameter $\operatorname{minrk}_\F^*(G)$. They construct a graph $G$ with $\theta(G)=9+7\sqrt{5}<28=\ham(G;\F)$ for every field $\F$, yet $\subbound(G;\F_{11})\leq\operatorname{minrk}_{\F_{11}}^*(G)\leq 24.5<9+7\sqrt{5}$.

\item For a graph $G$, is $\subbound(G)=\lim_{n\to\infty}\ham(G^{\boxtimes n})^{1/n}$? Corollary~\ref{cor:beatslimit} shows that $\subbound(G)\leq\ham(G^{\boxtimes n})^{1/n}$ for every positive integer $n$, so it is only necessary to verify the reverse inequality.
\end{itemize}

\bibliographystyle{abbrv}
\bibliography{references}

\end{document}